\documentclass[aps,pre,twocolumn,showpacs,superscriptaddress]{revtex4-1}

\usepackage{natbib} %[sort,compress]
\usepackage{longtable}
\usepackage{enumerate}
\usepackage{amsmath,amssymb}
\usepackage{color}
\usepackage{multirow}
\usepackage{graphicx}

\newcommand{\fracd}[2]{
\displaystyle{
\frac{ \displaystyle{#1} }{ \displaystyle{#2} }
}
}

\newcommand{\diff}[1]{\,{\rm d}{#1}}
\newcommand{\diag}{{\rm diag}}

\newcommand{\taub}{\tau_{\rm bulk}}

\newcommand{\ii}{{\dot{\imath}}}
\newcommand{\jj}{{\dot{\jmath}}}

\renewcommand{\c}{{\bf c}}
\newcommand{\ci}{\c_{\ii}}

\newcommand{\vbf}[1]{\mathbf{#1}}
\newcommand{\dt}{\Delta t} 
\newcommand{\dx}{\Delta x}
\newcommand{\rhor}{{\rho\!_{_{\circ}}}}

\newcommand{\acc}{g}
\newcommand{\bfacc}{\vbf{\acc}}

\newcommand{\dc}{\Delta_{\min}}

\newcommand{\uvel}{\vbf{u}}

\newcommand{\uj}{u_{\aj}}

\newcommand{\us}{{u_{\rm s}}}

\newcommand{\x}{\vbf{x}}
\renewcommand{\xi}{x_{\ai}}
\newcommand{\xj}{x_{\aj}}
\newcommand{\xk}{x_{\ak}}

\renewcommand{\Xi}{X_{\ai}}
\newcommand{\Xj}{X_{\aj}}

\newcommand{\Xs}{X_{\mc{S}}}
\newcommand{\Xc}{X_{\mc{C}}}

\newcommand{\ai}{1}
\newcommand{\aj}{2}
\newcommand{\ak}{3}

\newcommand{\e}{\vbf{e}}

\newcommand{\ej}{\e_{\aj}}

\newcommand{\dpl}{\avrg{(\nabla p)_{\xj}}}

\newcommand{\f}{f}

\newcommand{\fii}{\f_{\ii}}
\newcommand{\fq}{\f^{\rm eq}}
\newcommand{\fqi}{\fq_{\ii}}

\newcommand{\Vf}{V_{\rm f}}
\newcommand{\Vt}{V_{\rm t}}

\newcommand{\wwi}{\omega_{\ii}}
\newcommand{\cs}{{c_{\rm s}}}

\newcommand{\FEM}{\ensuremath{\text{FEM}}}
\newcommand{\LB}{\ensuremath{\text{LB}}}
\newcommand{\BGK}{\ensuremath{\text{BGK}}}

\newcommand{\MRT}{\ensuremath{\text{MRT}}}
\newcommand{\LBMRT}{\ensuremath{\text{LB-MRT}}}
\newcommand{\TRT}{\ensuremath{\text{TRT}}}

\newcommand{\lr}{\ensuremath{\text{L}}}
\newcommand{\ir}{\ensuremath{\text{M}}}
\newcommand{\hr}{\ensuremath{\text{H}}}

\renewcommand{\Re}{\ensuremath{\text{Re}}}
\newcommand{\Ma}{\ensuremath{\rm Ma}}

\newcommand{\mc}[1]{\mathcal{#1}}

\newcommand{\pwrr}[1]{10\sp{#1}}
\newcommand{\avrg}[1]{\langle #1 \rangle}
\newcommand{\avrgu}{\avrg{u}}

\newcommand{\m}{m}
\newcommand{\mi}{\m_{\ii}}
\newcommand{\bfm}{\vbf{\m}}

\newcommand{\mqi}{\m^{\mathrm{eq}}_{\ii}}
\newcommand{\bfmq}{{\bfm}^{\mathrm{eq}}}

\newcommand{\bfM}{\vbf{M}}

\newcommand{\bfhS}{\vbf{\check{S}}}
\newcommand{\hS}{\check{S}}

\begin{document}

\title{From creeping to inertial flow in porous media: a lattice Boltzmann -
  Finite Element study}

%Order to be determined:
\author{Ariel Narv\'{a}ez} 
\affiliation{ Department of Applied
   Physics, Eindhoven University of Technology, P.O. Box 513, NL-5600MB Eindhoven, The 
   Netherlands }

\author{Kazem Yazdchi} 
\affiliation{ Department of Engineering Technology,
  University of Twente, P.O. Box 217, NL-7500AE Enschede, The
  Netherlands }

\author{Stefan Luding} 
\affiliation{ Department of Engineering Technology,
  University of Twente, P.O. Box 217, NL-7500AE Enschede, The
  Netherlands }

\author{Jens Harting} 
\affiliation{ Department of Applied
  Physics, Eindhoven University of Technology, P.O. Box 513, NL-5600MB Eindhoven, The 
  Netherlands }
\affiliation{ Institute for Computational Physics, University of
  Stuttgart, Pfaffenwaldring 27, D-70569 Stuttgart, Germany }

% a.e.narvaez.salazar@tue.nl
% k.yazdchi@utwente.nl
% jharting@tue.nl

\date{\today}

\begin{abstract}
  The lattice Boltzmann method has been successfully applied for the
  simulation of flow through porous media in the creeping regime. Its
  technical properties, namely discretization, straightforward implementation
  and parallelization, are responsible for its popularity.  However, flow
  through porous media is not restricted to near zero Reynolds numbers since
  inertial effects play a role in numerous natural and industrial processes.
  In this paper we investigate the capability of the lattice Boltzmann method
  to correctly describe flow in porous media at moderate Reynolds numbers. The
  selection of the lattice resolution, the collision kernel and the boundary
  conditions becomes increasingly important and the challenge is to keep
  artifacts due to compressibility effects at a minimum.  The lattice
  Boltzmann results show an accurate quantitative agreement with Finite
  Element Method results and evidence the capability of the method to
  reproduce Darcy's law at low Reynolds numbers and Forchheimer's law at high
  Reynolds numbers.
\end{abstract}


\maketitle

\section{Introduction}
Predicting the transport properties of porous media, like the fluid
permeability, defined by Darcy's law, or heat conductivity, is of paramount
importance in chemical, mechanical, geological, environmental and petroleum
industries. Flow situations in porous media are not restricted to the creeping
flow regime, i.e., near zero Reynolds numbers where Darcy's law
applies~\cite{Bear:1972}. Many important natural and industrial processes are
characterized by large Reynolds numbers, where inertial effects also play a
role. Examples include gas flow through a catalytic converter, groundwater
flow, filtration processes and the flow of air in our
lungs~\cite{MPT_2009}. Hence, accurate simulation methods are needed to
improve our understanding of these processes.

The lattice Boltzmann (\LB) method has become an efficient
tool~\cite{bib:ferreol-rothman,1990PhFl.2.2085C,bib:qian-dhumieres-lallemand}
as an alternative to a direct numerical solution of the Navier-Stokes
equation~\cite{PhysRevB.46.6080,MAKHT02} for simulating fluid flow in complex
geometries such as porous media. Historically, the
\LB\ method was developed from the lattice gas
automata~\cite{bib:qian-dhumieres-lallemand,bib:succi-01}, replacing the
number of particles in each lattice direction with the ensemble average of the
single particle distribution function, and the discrete collision rule with a
linear collision operator. In the \LB\ method all computations involve local
variables so that it can be parallelized easily~\cite{MAKHT02}. 
Together with uniform grids and thus straightforward discretization, the \LB\
method has become very popular in the field of porous media flow
simulations. With the advent of more powerful computers it became possible to
perform detailed simulations of finely resolved complex samples~\cite{bib:aharonov-rothman,bib:auzeraisGRL96,bib:ferreol-rothman,1990PhFl.2.2085C,KL97,HKL01,bib:cf.CPaLLuCMi.2006,NZRHH10,NH10,MAKHT02}.

In this work we investigate the accuracy of the \LB\ method in flow regimes
beyond Darcy's regime. We focus on limitations of the method with respect to
the lattice resolution and the selection of parameters. In particular, we
address the requirement of reducing undesired compressibility effects by
keeping the Mach number low, and how this influences the achievable maximum
Reynolds numbers. In order to keep the compressibility as moderate as possible
a new simulation setup combining an injection channel, density boundary
conditions, and an external body force for driving the fluid is introduced.
Nevertheless, for high Reynolds numbers, a low fluid viscosity and high
resolution are required. Thus, a thorough investigation of the impact of
the compressibility on the measured permeability is presented.

To confirm the validity of the \LB\ results, they are compared to Finite
Element Method (\FEM) simulations that are performed with the commercial
software package ANSYS. During the past decades, the \FEM\ has been widely
used to simulate fluid flow through porous media. It is known that \FEM\ can
deal with complex pore geometries and boundary conditions, see for examples
Refs.~\cite{Zabaras, girault, thomasset}. Tezduyar et al.~\cite{Tezduyar} have
developed the so-called deforming-spatial-domain/space-time (DSD/ST) procedure
for flow problems with deforming interfaces using the so-called Arbitrary
Lagrangian-Eulerian (ALE) methods and a space-time finite element method. This
approach was based on fully resolved simulations around particles and
therefore computationally expensive in dense flows. For an overview of some
finite element and finite difference techniques for incompressible fluid flow
see Ref.~\cite{Fletcher} and for the efficiency of the solution
algorithms Ref.~\cite{Turek}. In recent studies, Yazdchi et al. relate the macroscopic properties of porous media, namely permeability and inertial coefficients, to their microstructure and porosity~\cite{kazemluding,kazemluding2}.

The remainder of this article is organised as follows: In section II, an
introduction to porous media flow including laminar and weakly inertial flow,
i.e. creeping flow, named Darcy's flow, and Forchheimer's flow is given,
respectively. In section III, the \LB\ and \FEM\ methods are introduced along
with the description of the simulation setup.  In section IV, we demonstrate
quantitative agreement of \LB\ and \FEM\ simulations with theoretical
predictions for fluid flow in porous media in the above mentioned
regimes. Finally we analyze and compare the results before we conclude.

%Further analysis shows a good agreement
%with theory, i.e. Darcy's
%law at low Reynolds numbers and Forchheimer's law at high Reynolds numbers,
%and also confirms the presence of the well documented transition regime. The
%data analysis shows Darcy's regime for $\Re<10^{0}$, a transient regime,
%described by Darcy's law plus a cubic correction of the surface velocity, for
%$10^{0}<\Re<10^{1}$, and Forchheimer's regime for higher Reynolds numbers.
%

\section{Porous media flow} 
Weak inertial flow, also named creeping flow (near zero Reynolds numbers) in
porous media, can be described by Darcy's law~\cite{bib:darcy,Dullien:1992}.
It is defined by
\begin{equation}
\label{darcy:eq}
\kappa = - \mu \frac{\us}{\dpl},
\end{equation}
where the porous medium parameter $\kappa$ (always positive) is named the permeability of the medium. $\dpl$ represents the average pressure gradient in the
direction of the flow $\xj$, see Fig.~\ref{domain:fig}, and $\mu$
represents the dynamic viscosity of the fluid. $\us$ represents the
superficial velocity defined by
\begin{equation}
\label{superficial:vel:eq}
\us = \frac{1}{\Vt}\int_{\Vf} \uj \diff{V},
\end{equation}
where $\uj$ is the fluid velocity in the flow direction. $V_{\rm t}$ and
$V_{\rm f}$ represent the total volume and the fluid volume, respectively.
The average velocity within the porous medium $\avrgu$ is related to the
superficial velocity by $\us=\varepsilon\avrgu$, where $\varepsilon$ is the
porosity of the medium. According to Eq.~\eqref{darcy:eq}, Darcy's law
corresponds to a linear relation between the average pressure gradient $\dpl$
and the superficial velocity $\us$, in the literature also referred to as
seepage velocity~\cite{MA91}. Darcy's law was
obtained empirically in 1846, but it can be derived from the continuous moment
and mass balance assuming that the solid-fluid hydrodynamic interaction is
proportional to the relative solid-fluid velocity~\cite{Mass1984}.

When the Reynolds number is increased, inertia effects become relevant
($\Re \approx \mathcal{O}(1)$). The average pressure gradient $\dpl$
and the superficial velocity $\us$ do not follow a linear relation anymore as
it was empirically shown by Forchheimer~\cite{Forchh01,ACAMS_1999}. For this
flow regime, named Forchheimer's regime, a quadratic term in $\us$ is included
to take into account the inertia effects.  According to
Massarani~\cite{Mass1984}, Forchheimer's law can be written as
\begin{equation}
\label{forchheimer:eq}
\dpl = -\left( \frac{\mu}{\kappa} +
\frac{\hat\rho\!_{\circ} c}{\sqrt{\kappa}}\us\right)\us,
\end{equation}
where $c$ is a positive dimensionless parameter and $\hat\rho\!_{\circ}$ is
the reference fluid density. The quadratic term in Eq.~\eqref{forchheimer:eq}
describes a linear relationship between the solid-fluid hydrodynamic
interaction and the relative solid-fluid velocity. Darcy's law is recovered in
the limit of $\us\rightarrow$0.
% and Forchheimer's law is commonly considered to be
%valid in the creeping regime.

A transition from creeping flow to inertial flow has been reported in many
studies, see for example the work of Koch and Ladd on flow in a random
array of cylinders~\cite{KL97} and spheres~\cite{HKL01}. Also the
existence of a transition regime between the creeping and inertial regimes
has been reported in the past~\cite{SA99}. This departure from Darcy's
regime is of the order of $\us^{3}$~\cite{kazemluding,AGO07,WL91,MA91,FGQ97} and the
relation between $\dpl$ and $\us$ in this short interval is then given by
\begin{equation}
\label{cubic:eq}
\dpl = -\left( \frac{\mu}{\kappa} + \frac{\gamma {\hat\rho\!_{\circ}}^{2}}{\mu} \us^{2} 
\right) \us,
\end{equation}
where the cubic term in $\us$ with positive parameter $\gamma$ represents
the weak inertia correction.

\section{Simulation methods}
\subsection{The lattice Boltzmann Method}
The integration of the Boltzmann equation on a regular lattice using a
discrete set of velocities $\ci$ defines the lattice Boltmann method. The
lattice is defined by the spacing $\dx$, with the discrete velocity in units of $\dx/\dt$,
where $\dt$ represents the timestep. Thus, the basic differential equation for
the method is
\begin{equation}
\label{LB:eq}
\fii(\x+\dt\,\ci,t+\dt)-\fii(\x,t) = 
-\frac{\dt}{\tau}\left(\fii(\x,t)-\fqi(\x,t)\right),
\end{equation}
where $\fii(\x,t)$ represents the number of particles moving at position
$\x$ with velocity $\ci$ at discrete time $t$. The term on the right hand side
represents the collision operator as introduced by Bhatnagar, Gross and
Krook (\BGK)~\cite{bib:chen-chen-martinez-matthaeus,bib:bgk}, which
approximates the collision through a linearization towards the
equilibrium distribution function $\fqi(\x,t)$ with a unique
relaxation time $\tau$. This value is restricted to $\tau/\dt > 1/2$
assuring positive viscosity. If $\tau/\dt$ approaches $1/2$
numerical instabilities can arise~\cite{Sterling:1996:SAL:227647.227666}.
This collision operator is often referred to as ``single relaxation time'' or
\BGK\ model. Under the assumption of very low Knudsen and
Mach numbers (\Ma) which assures small compressibility effects,
$\fqi(\x,t)$ is calculated by a second order Taylor expansion of the
Maxwell distribution as~\cite{bib:qian-dhumieres-lallemand}
\begin{equation}
\label{eq:dist:eq}
\fqi(\x,t) = \wwi \frac{\rho}{\rhor} 
\left( 1 + \frac{\uvel\cdot\ci}{{\cs}^{2}} 
+ \frac{ (\uvel\cdot\ci)^{2}}{2{\cs}^{4}}
- \frac{\uvel\cdot\uvel}{2{\cs}^{2}} \right).
\end{equation}
$\rhor$ is a reference density.
The coefficients $\wwi$ are called lattice weights and are chosen to assure
conservation of mass and momentum. They differ with lattice type, number of
space dimensions and number of discrete velocities.  For the cubic fashion
3D~lattice with 19 discrete velocities ($i=0,..,18$) used in this work they
are $1/3$, $1/18$ and $1/36$ for the rest particles, the particles moving
parallel to the $\xi$, $\xj$, or $\xk$, and the particles moving in diagonal
directions, respectively~\cite{bib:qian, bib:qian-dhumieres-lallemand}.  Even
though the system of interest in this paper is intrinsically two-dimensional,
we apply a three-dimensional implementation of the \LB\ method and compute the
flow in a very flat simulation domain with periodic boundary conditions in the
$\xk$ direction. We do not expect this to have any influence on the simulation
results, but it allows us to use our well tested implementation ``LB3D''. The
only disadvantage are higher computational costs, but we do not report on the
amount of CPU time required for a given flow problem in this paper.

No-slip boundary conditions on the solid walls are implemented by mid-plane
bounce back rules~\cite{SukopThorne2007}.  The \BGK\ model is known to suffer
from an artificial viscosity dependent slip at boundaries if these boundary
conditions are used.  An alternative approach for the collision operator,
which reduces this well known drawback of the \BGK\ model, is the multi
relaxation time (\MRT) method~\cite{bib:cf.CPaLLuCMi.2006,NZRHH10}. Here, the
right hand side of Eq.~\eqref{LB:eq} is replaced by the expression
\begin{equation}
\label{mrt:collision:eq}
-\dt \left[{\bfM}^{-1} \cdot \bfhS \cdot \left( \bfm(\x,t) -
 \bfmq(\x,t) \right) \right]_{\ii},
\end{equation}
where $\bfM$ is a linear transformation chosen such that the moments
\begin{equation}
\mi (\x,t)=\sum_{\jj} M_{\ii\,\jj} \, \f_{\jj}(\x,t)
\end{equation}
represent hydrodynamic modes of the problem. We use the definitions
given in~\cite{2002RSPTA.360..437D}, where $\m_{0}$ is the fluid
density, $\m_{2}$ represents the kinetic energy, $\mi$ with $\ii=3,5,7$ is the
momentum flux and $\mi$, with $\ii=9,11,13,14,15$ are components of
the symmetric traceless viscous stress tensor. During the collision step the
density and the momentum flux are conserved so that $\mi=\mqi$ with
$i=0,3,5,7$. The non-conserved equilibrium moments $\mqi$, $i\neq
0,3,5,7$, are assumed to be functions of these conserved moments and
explicitly given in~\cite{2002RSPTA.360..437D}.  $\bfhS$ is a diagonal
matrix $\hS_{\ii\,\jj} = \check{s}_{\ii} \, \delta_{\ii\,\jj}$. The
diagonal elements $\tau_{\ii}=1/\check{s}_{\ii}$ in the collision
matrix are the relaxation time of the moment $\mi$. One has
$\check{s}_{0}=\check{s}_{3}=\check{s}_{5}=\check{s}_{7}=0$, because
the corresponding moments are conserved. $\check{s}_{1}=1/\taub$
describes the relaxation of the energy and
$\check{s}_{9}=\check{s}_{11}=\check{s}_{13}=\check{s}_{14}=\check{s}_{15}=1/
\tau$ the relaxation of the stress tensor components. The remaining
diagonal elements of $\bfhS$ are chosen such that one has
\begin{equation}
\begin{split}
\bfhS = \diag & (0,1/\taub,1.4,0,1.2,0,1.2,0,1.2,1/\tau,\\
& 1.4,1/\tau,1.4,1/\tau,1/\tau,1/\tau,1.98,1.98,1.98),
\end{split}
\end{equation}
to optimize the algorithm
performance~\cite{2000PhRvE..61.6546L,2002RSPTA.360..437D}. The two
relaxation times $\tau$ and $\taub$, are restricted as well as for the \BGK\
method to be $>\dt/2$, but allow to define the kinematic and bulk viscosity,
respectively.  This multi relaxation time scheme is commonly referred to as ``two
relaxation time'' (\TRT) method. An alternative \TRT\ implementation can be
found in~\cite{el00178,el00173}.

The macroscopic density $\rho(\x,t)$ and velocity $\uvel(\x,t)$ are
obtained from $\fii(\x,t)$ as
\begin{eqnarray}
\label{rho:eq}
\rho(\x,t) & = & \rhor \sum_{\ii} \fii(\x,t), \\
\label{uvel:eq}
\uvel(\x,t) & = & 
\frac{\rhor}{\rho(\x,t)} \sum_{\ii}\fii(\x,t) \, \ci.
\end{eqnarray}
The pressure is given by
\begin{equation}
\label{p:eq}
p(\x,t) = {\cs}^{2} \, \rho(\x,t), 
\end{equation}
where $\cs = 1/\sqrt{3}(\dx/\dt)$ is the speed of
sound~\cite{bib:qian-dhumieres-lallemand,bib:succi-01}. The kinematic
viscosity of the fluid $\nu=\mu/\rho$ is a function of the discretization parameters,
$\dx$ and $\dt$, and the relaxation time
$\tau$~\cite{bib:chapman-cowling,Wolf05}. It is given by
\begin{equation}
\label{nu:eq}
\nu = \frac{\cs^2 \, \dt}{2} \left( 2\frac{\tau}{\dt} - 1 \right).
\end{equation}
%
%The reference density is set to $\rhor = \pwrr{3}\,
%\mathrm{kg}\,\mathrm{m}^{-3}$ and the bulk relaxation time is kept fixed at
%$\taub/\dt = 0.84$~\cite{2002RSPTA.360..437D}. 

\subsection{The Finite Element Method} 
The velocity and pressure profiles through the system can be obtained
from the solution of the conservation laws, namely, the continuity
equation (conservation of mass) and the Navier-Stokes equations
(conservation of momentum). In the absence of body forces,
but assuming a constant density (i.e. incompressible flow) and steady state
flow conditions, the equations of conservation of mass and momentum for a Newtonian fluid are
simplified to
\begin{eqnarray}
\label{fem1:eq}
  \nabla\cdot\uvel & = & 0, \\
\label{fem2:eq}
  \rho( \uvel\cdot\nabla \uvel ) & = & -\nabla p + \mu \nabla^{2}\uvel.
\end{eqnarray}
The \FEM\ makes use of the variational formulations that allow the
transformation of the above equations into a system of linear algebraic
equations, which can be solved using a simple LU decomposition or iterative
algorithms~\cite{Kandhai_1999,saurabh_2012}. Stable discretizations of the
above equations are difficult to construct and it is known that the
incompressibility constraint is not strongly enforced when using continuous
polynomial shape functions for pressure. See~\cite{Turek} for a detailed
theory and discussion. Langtangen et al. present an
overview of the most common numerical solution strategies, including fully
implicit formulations, artificial compressibility methods, penalty
formulations and operator splitting methods~\cite{Langtangen}. Using the conventional \FEM\
scheme, we solve the above equations with the commercial software
ANSYS~\cite{bib:Yazdchia}. The nonlinear solution procedure used in ANSYS
belongs to a general class of Semi-Implicit Methods for Pressure Linked
Equations (SIMPLE). On the flow domain, the steady state Navier-Stokes
equations combined with the continuity equations are discretized into linear
triangular elements.  They are then solved using a segregated sequential
solution algorithm. This means that element matrices are formed, assembled and
the resulting system is solved using the Gaussian elimination algorithm for
each degree of freedom separately. The number of iterations required to
achieve a converged solution may vary considerably, depending on the number of
elements, inertial contribution and the stability of the problem. Some more
technical details are given in~\cite{bib:Yazdchia}.

By knowing the fluid velocity field, the superficial velocity is then
calculated from Eq.~\eqref{superficial:vel:eq}. Recently, using \FEM\
simulations, Yazdchi and Luding~\cite{kazemluding} show that for both ordered and random fibre arrays, the weak inertia
correction to the linear Darcy relation is third power in superficial
velocity, up to small Re $\approx$ 1-5. When attempting to fit the data with a
particularly simple relation, a non-integer power law performs astonishingly
well up to the moderate Re $\approx$ 30.

A typical unstructured, fine
and triangular \FEM\ mesh is shown in Fig.~\ref{domain:fig}.  The mesh size effect is examined by comparing the simulation
results for different resolutions. The mesh refinement is done on element level, meaning that a finer grid is overlaid on the coarse one. In order to be
able to apply periodic boundary conditions in $\xi$ direction, see
Fig.~\ref{domain:fig}, we discretize the system such that we come up with the
same number of nodes on the left and right boundary. Periodic boundary
conditions are applied by setting additional constrains, i.e. same velocity, on
these nodes.

\subsection{Computational Domain and Simulation Setup}
\begin{figure*}[t]
\includegraphics[width=0.6\textwidth]{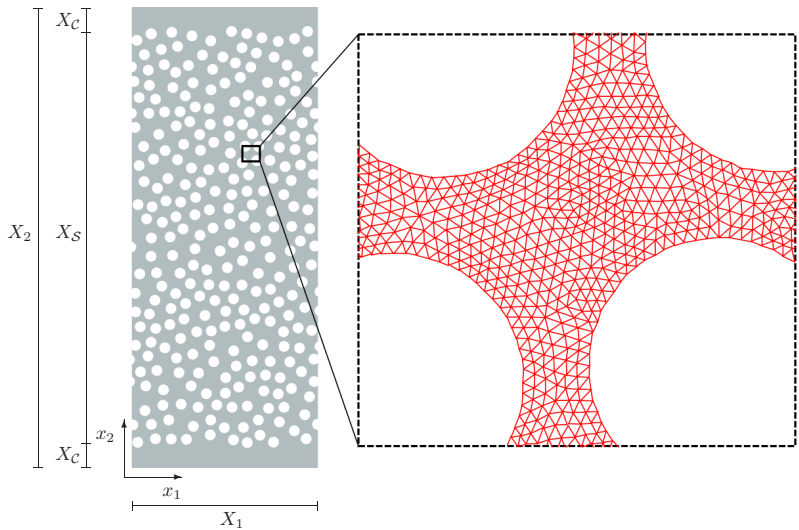} 
\caption{Computational domain with dimensions $\Xi\times\Xj$
  ($\Xi=0.4\,\Xj$). The sample is composed of 266~circles with radius
  $r$ representing 1.25\% of $\Xj$. They are randomly distributed
  assuring a minimum circle center distance $\dc$ of 3.125\% of
  $\Xj$. Pure fluids chambers with length of 5\% of $\Xj$ are placed after and
  before the sample. The right inset shows a zoom of a typical unstructured,
  fine and triangular \FEM\ mesh.}
\label{domain:fig}
\end{figure*}

The computational domain is presented in Fig.~\ref{domain:fig}.
%hasdimensions $\Xi\!\times\!\Xj$, where $\Xi=0.4\cdot\Xj$. Chambers are
%placed right before and after the sample to provide a fluid reservoir.
%Their length $\Xc$ represents 5\% of $\Xj$. 
The porous sample has
a length of $\Xs$ representing $0.9\,\Xj$, and it is composed of
266~circles, randomly distributed with a radius $r$ of 1.25\% of
$\Xj$.  A minimum distance $\dc$ of 3.125\% of $\Xj$ is imposed
between the circle centers. The sample porosity then follows to be $\varepsilon =
0.63730$.
% 1.-266*(1.25/100.0)^2*pi/(40.0/100.0*90.0/100.0)

For the \LB\ method three different resolutions are used, namely low
resolution (\lr), intermediate resolution (\ir), and high resolution
(\hr), see Table.~\ref{domain:lb:tbl} for details. For the \FEM\ method
also three resolutions are used, corresponding to 22048, 49670, and
982376~triangular elements, respectively. These meshes are referred to
using the same abbreviations as for the \LB\ simulations. However, one should
keep in mind that the number of discretisation elements used in both
methods cannot be compared easily since the \LB\ simulations utilize a regular
Cartesian lattice, while the \FEM\ simulations are based on unstructured grids
with locally varying resolution.

%awk < domain128.txt '{ if ($4==0) count+=1} END {print count/4}'
%awk < domain256.txt '{ if ($4==0) count+=1} END {print count/4}'
%awk < domain512.txt '{ if ($4==0) count+=1} END {print count/4}'
%awk < domain1024.txt '{ if ($4==0) count+=1} END {print count/4}'
%
\begin{table*}[ht]
  \caption{Domain discretization for the \LB\ simulations.}
\label{domain:lb:tbl} 
\begin{tabular}{| c | c c c c c c |}
\hline
%& $\dx$ &  $\dt$  & & & & & &  & Number of \\
& & & & & &  \\
\raisebox{1.5ex}[-1.5ex]{Resolution} & 
%[$\pwrr{-6}\mathrm{m}]$ & 
%[$\pwrr{-6}\,\mathrm{s}]$ & 
\raisebox{1.5ex}[-1.5ex]{$\fracd{\Xi}{\dx}$} & 
\raisebox{1.5ex}[-1.5ex]{$\fracd{\Xj}{\dx}$} & 
\raisebox{1.5ex}[-1.5ex]{$\fracd{\Xs}{\dx}$} & 
\raisebox{1.5ex}[-1.5ex]{$\fracd{\Xc}{\dx}$} & 
\raisebox{1.5ex}[-1.5ex]{$\fracd{r}{\dx}$} & 
\raisebox{1.5ex}[-1.5ex]{$\fracd{\dc}{\dx}$} 
%& elements 
\\
\hline
%low (\lr)          & 2.00 & 2.00 &  128 &  320 &  288 &  16 &  4 & 10 \\
% &  22048 \\
%medium (\ir)       & 1.00 & 1.00 &  256 &  640 &  576 &  32 &  8 & 20 \\
% &  110362 \\
%high (\hr)         & 0.50 & 0.50 &  512 & 1280 & 1152 &  64 & 16 & 40 \\
%&  441414 \\
low (\lr)          &  128 &  320 &  288 &  16 &  4 & 10 \\
medium (\ir)       &  256 &  640 &  576 &  32 &  8 & 20 \\
high (\hr)         &  512 & 1280 & 1152 &  64 & 16 & 40 \\
\hline 
\end{tabular}
% \begin{tabular}{| c | c c c c c c c c c | c c |}
% \cline{2-12}
% \multicolumn{1}{c}{} & \multicolumn{9}{|c|}{\LB} & \multicolumn{2}{|c|}{\FEM} \\
% \hline
% & $\dx$ &  $\dt$  & & & & & &  & Number of & $r$ & Number of\\
% \raisebox{1.5ex}[-1.5ex]{Resolution} & 
% [$\pwrr{-6}\mathrm{m}]$ & 
% [$\pwrr{-6}\,\mathrm{s}]$ & 
% \raisebox{1.5ex}[-1.5ex]{$\fracd{\Xi}{\dx}$} & 
% \raisebox{1.5ex}[-1.5ex]{$\fracd{\Xj}{\dx}$} & 
% \raisebox{1.5ex}[-1.5ex]{$\fracd{\Xs}{\dx}$} & 
% \raisebox{1.5ex}[-1.5ex]{$\fracd{\Xc}{\dx}$} & 
% \raisebox{1.5ex}[-1.5ex]{$\fracd{r}{\dx}$} & 
% \raisebox{1.5ex}[-1.5ex]{$\fracd{\dc}{\dx}$} & 
% elements & 
% [$\pwrr{-6}\,\mathrm{m}]$ & 
% elements\\
% \hline
% low (L)          & 2.00 & 2.00 &  128 &  320 &  288 &  16 &  4 & 10 &   27583 & 8 &  22048 \\
% medium (M)       & 1.00 & 1.00 &  256 &  640 &  576 &  32 &  8 & 20 &  110362 & 8 &  49670 \\
% high (H)         & 0.50 & 0.50 &  512 & 1280 & 1152 &  64 & 16 & 40 &  441414 & 8 & 353088 \\
% super (X)        & 0.25 & 0.25 & 1024 & 2560 & 2304 & 128 & 32 & 80 & 1765823 & 8 & 982376 \\
% \hline 
% \end{tabular}
\end{table*}

In the \LB\ simulations a constant acceleration $\bfacc=\acc\ej$ drives the fluid using Guo's
method~\cite{Guo02}. This acceleration defines the pressure gradient and acts
on the fluid inside the sample, i.e. from $\xj=\Xc$ to $\xj=\Xc+\Xs$.
Together with this, on-site pressure boundary conditions are implemented on
the inlet-plane $\xj=0$ and on the outlet-plane $\xj=\Xj$ setting a constant
density value $\rhor$ (reference density, i.e. $\sum \fii = 1$) on both
planes. See Eq.~\eqref{p:eq} for the relation between the density and pressure
within the \LB\ method. These pressure boundary conditions are implemented
using the method of Zou \& He~\cite{bib:pf.QZoXHe.1997,HH08b}.  Different
values of the acceleration $\acc$ are used in order to study different \Re\
regimes.  In our earlier work we proposed to use just on-site pressure
boundary conditions at the inlet- and outlet-planes, ($\sum \fii = 1+\delta $
and $\sum \fii = 1-\delta $, respectively) to impose a pressure drop of
$2\delta$~\cite{NH10}. The here presented new setup allows to reduce
compressibility effects inside the sample when the Reynolds number is increased.
Periodic boundary conditions are applied in the $\xi$ direction. 

In the \FEM\ simulations, a pressure drop is imposed while the density is kept
constant to drive the fluid. Further, we impose zero velocity on the surface of
the fibres and also apply periodic boundary conditions in the $\xi$ direction.

\section{Simulation Results} 
We present a detailed analysis of the simulation of porous media flow at low to
moderate Reynolds numbers using the \LB\ method. Special attention is given to its
collision kernel, the discretization and the value of the relaxation time.  To
increase \Re\ either the average flow velocity can be increased or the fluid
viscosity can be decreased. However, the range of these parameters is limited:
the lattice Boltzmann method is known to reproduce the Navier-Stokes equations
in the low Mach number limit only, i.e., at high flow velocities compressibility
artifacts can occur and render the results invalid. To validate the obtained
data and to understand the impact of these limits on the precision of the \LB\
results, we present a quantitative comparison with \FEM\ results and
theoretical predictions.

\begin{figure}[t]
\includegraphics[width=1.0\columnwidth]{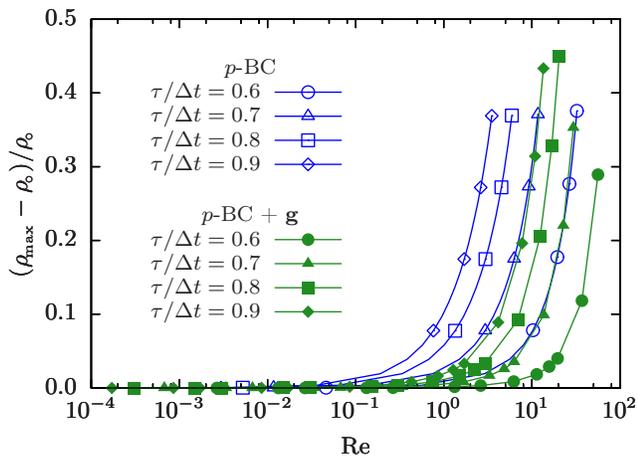}
\caption{Normalized maximum fluid density as a measure for the effect of
  compressibility versus Reynolds number. Open symbols denote data obtained
  from \LBMRT\ simulations where the fluid is driven by an imposed pressure
  drop ($p$-BC) as suggested in~\cite{NH10}. When combining a fixed density at
  the in- and outlet of the domain with a driving body force ($p$-BC +
  $\bfacc$), compressibility effects can be reduced substantially. Together
  with low values of the relaxation time $\tau/\dt$ higher \Re\ can be reached
  (solid symbols).}
\label{compr:rho:fig} 
\end{figure}

For the \LB\ simulations $100,000$ timesteps assure the steady state. The
superficial velocity defined by Eq.~\eqref{superficial:vel:eq} is calculated
from the steady state \LB\ data by
\begin{equation}
\label{us:lb:eq}
\us = \fracd{(\dx)^{2}}{\Xi\,\Xj} \sum_{\x\in\mc{X}} \uj(\x), 
\end{equation}
where $\mc{X}$ represents all fluid nodes in the simulation domain. 
The average pressure gradient is expected to be
\begin{equation}
\label{dpl:lb:eq}
\dpl = -g\rhor. 
\end{equation}
The Mach and Reynolds numbers are defined by
\begin{eqnarray}
\Ma & = & \frac{\avrgu}{\cs}, \\
\Re & = & \frac{\avrgu\,r}{\nu}.
\end{eqnarray}
Fig.~\ref{compr:rho:fig} shows the relative maximum fluid density as a measure
for the compressibility. The data are obtained from \LBMRT\ simulations of
flow in the model geometry introduced above. While the open symbols represent
data obtained using standard pressure drop boundary conditions
(see~\cite{NH10}), the closed symbols describe data obtained with the newly
proposed simulation setup combining pressure boundary conditions with a body
force $\bfacc$. In Fig.~\ref{compr:rho:fig} the compressibility and thus the
increase of the relative maximum density in the system increases for higher
\Re. Further, a clear reduction of compressibility effects is obtained when
the newly proposed simulation setup is used.

Fig.~\ref{compr:rho:fig} also shows the effect of a reduced relaxation time
(and thus viscosity) on the maximum reachable \Re\ and the related
compressibility effects. The compressibility of the fluid starts to become
less important for lower viscosities and the combination of a low relaxation
time ($\tau/\dt=0.6$) together with the combined pressure and body force
boundary conditions leads to higher maximum Reynolds numbers. With this
combination, \Re\ of the order of 30 can be reached before, at large \Re,
fluctuations become too large for reliable analysis of the simulations. 

\begin{figure}[h]
\includegraphics[width=1.0\columnwidth]{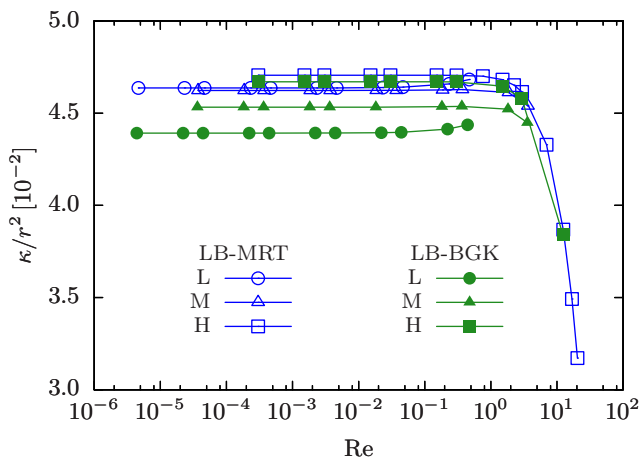}
\caption{Resolution and collision kernel ($\tau/\dt=0.6$) analysis for
  permeability estimation vs. \Re. The results show a dependency with the
  domain discretization, which is stronger when the \BGK\ collision kernel is
  used.
%The estimated permeability
%remains constant for small \Re, but for higher values Darcy's law does not hold
%since the permeability does not remain constant.
}
\label{k:fig} 
\end{figure}

Fig.~\ref{k:fig} shows the permeability estimate obtained from \LB\ simulations
using different discretizations and collision kernels. In our previous
work it was shown for \BGK\ simulations of 3D Poiseuille flow
that when a relaxation time close to $\tau/\dt=0.8$ is used, the dependency of
the results on the discretization is minimal~\cite{NZRHH10}. As it is also
demonstrated in the same article, finding such an optimal value for the
relaxation time is impossible for more realistic samples such as a discretized
Fontainebleau sandstone or the array of cylinders as used here: a strong
dependency on the discretization appears when the \BGK\ model is used, even
though the relaxation time is set to $\tau/\dt=0.8$. By using the \MRT\ model
this dependency can be decreased substantially. As can be observed in
Fig.~\ref{k:fig}, the results of both collision kernels are qualitatively
similar, i.e., both methods correctly reproduce the expected shape of the curve
for comparable maximum \Re. One can see that in the case of the low resolution
sample (\lr) the departure from the Darcy's regime (constant permeability) is
to higher permeability values, which is unphysical since $c$ in
Eq.~\eqref{forchheimer:eq} is a non-negative parameter. In the case of
intermediate (\ir) and high resolution (\hr), the departure is to lower
permeability values, but only using the high resolution sample it is possible
to simulate high-enough Reynolds numbers to analyze this phenomenon.

\begin{figure}[h]
\includegraphics[width=1.0\columnwidth]{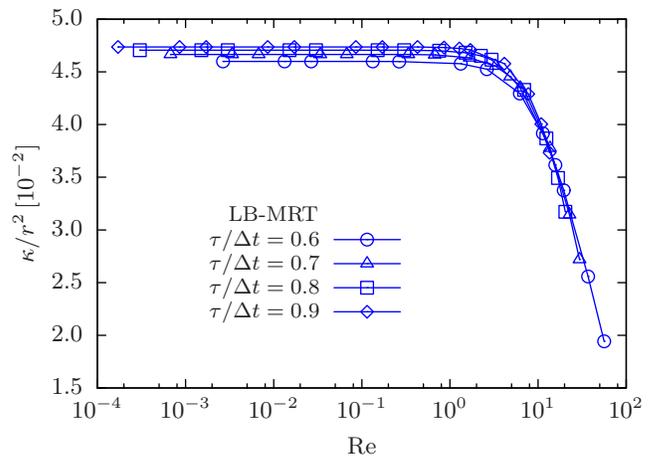}
\caption{Relaxation time analysis for permeability estimation vs. \Re\
  calculated using the \LBMRT\ method and the high resolution (\hr) computational
  domain. Different values for the relaxation time $\tau/\dt$ were used. The
  results only differ by $\approx 3\%$ and the highest \Re\ can be reached
  with the smallest value of $\tau/\dt=0.6$.}
\label{k:hr:fig}
\end{figure}

Fig.~\ref{k:hr:fig} shows the permeability estimation for the \LBMRT\ method
using different values for the relaxation time. Due the resolution used for
this study (\hr) together with the MRT collision kernel it is not surprising
that the results do not differ much quantitatively, but only by about 3\%. The
main difference is in the highest possible Reynolds numbers which can be
reached by using a small value of the relaxation time $\tau/\dt=0.6$, in
accordance with Fig.~\ref{compr:rho:fig}. As mentioned above numerical
instability arises when the relaxation time approaches $\tau/\dt=0.5$. For
this reason the smallest value used in this work is $\tau/\dt=0.6$.

\begin{figure}[h]
\includegraphics[width=1.0\columnwidth]{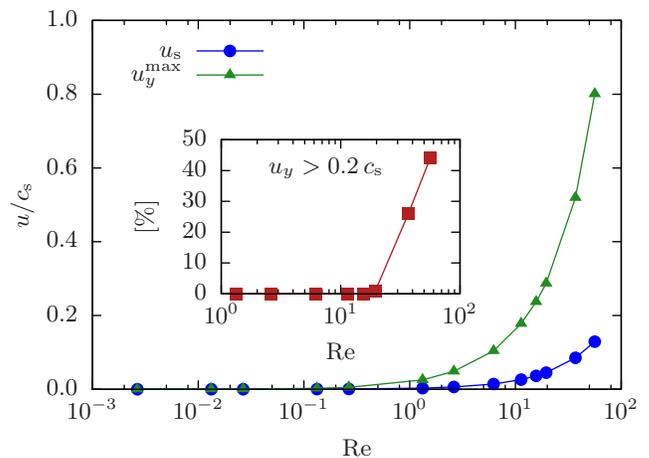}
\caption{Superficial velocity $\us$ and maximal velocity
  $u_{y}^{\max}$ inside the sample for \LBMRT\ and $\tau/\dt=0.6$. The
  inset shows the percentage of fluid nodes
  with velocity $u_{y}$ higher than 20\% of the speed of sound.}
\label{us:umax:fig}
\end{figure}

For porous media flow calculations using \LBMRT\ and $\tau/\dt=0.6$,
Fig.~\ref{compr:rho:fig} shows that at high Reynolds numbers the
compressibility is of the order of 30\%. Furthermore as we can see in
Fig.~\ref{us:umax:fig} the superficial velocity remains low enough to keep
Mach numbers below $\approx \pwrr{-1}$. However, inside the sample there are
zones with high velocity. Indeed, for the two higher Reynolds numbers
simulated, on $\approx25\%$ and $\approx45\%$ of the fluid nodes the velocity
is higher than 20\% of the speed of sound $\cs$, see
Fig.~\ref{us:umax:fig}. Such high velocities and strong compressibility which
are present at high Reynolds numbers highly question the validity of the
results.

\begin{figure}[h]
\includegraphics[width=1.0\columnwidth]{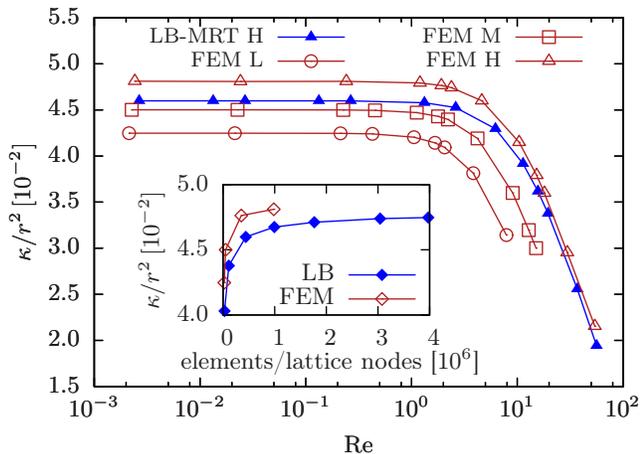}
\caption{Comparison of the permeability vs. \Re\ obtained using the \LBMRT\
  ($\tau/\dt=0.6$ and \hr) and different resolutions of the \FEM\ grid. The
  inset shows the resolution dependent permeability for \LB\ (\LBMRT\ and
  $\tau/\dt=0.6$) and \FEM\ simulations at $\Re\approx 10^{-3}$.
\label{k:LB:FEM:fig}
}
\end{figure}

\begin{figure}[h]
\includegraphics[width=1.0\columnwidth]{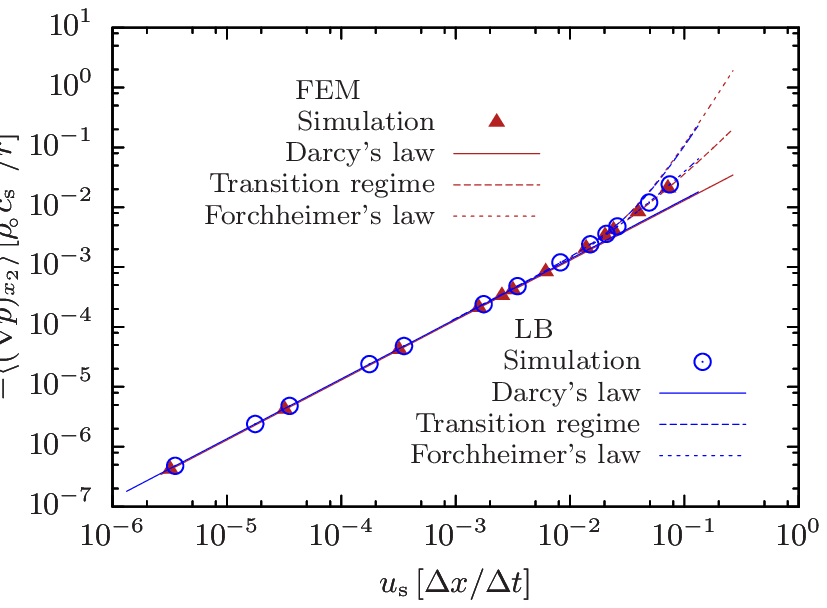}
\includegraphics[width=1.0\columnwidth]{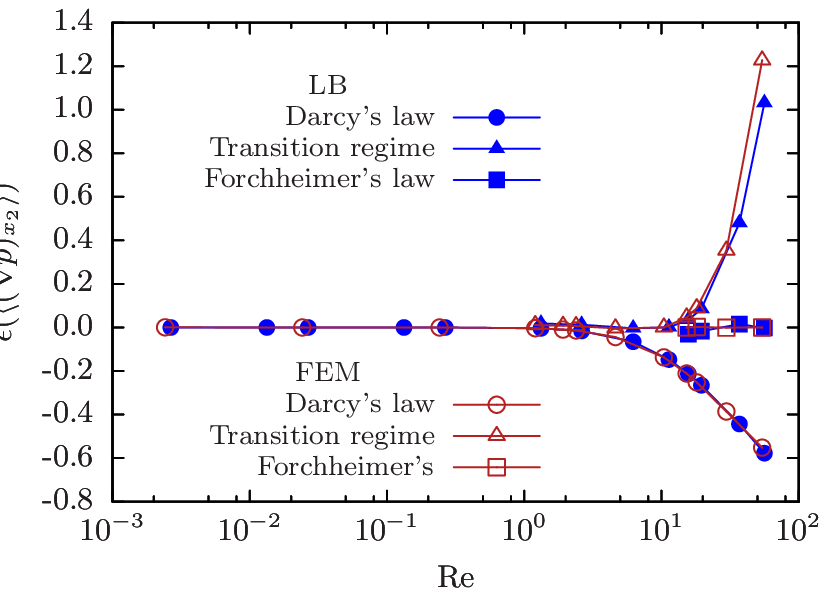}
\caption{(Top) Fit of the expressions which relate the superficial velocity
  $\us$ and the average pressure gradient $\dpl$ for the porous media
  flow. (Bottom) Relative error of the fit presented above. The relative error
  shows that Darcy's law fits the simulation results for $\Re \lesssim
  \pwrr{0}$. Above this value the fit of the expressions given by
  Eq.~\eqref{cubic:eq} is better than the fit with Darcy's law.
%For higher values of \Re, $\Re \gtrsim \pwrr{1}$, Forchheimer's law fits
%  the results best. Our \LB\ and \FEM\ simulations show very good
%agreement and confirm the presence of a transition regime.
}
\label{fit:fig} 
\end{figure}

To investigate the validity of the results \FEM\ simulations are performed as a
benchmark for the \LB\ simulations. In Fig.~\ref{k:LB:FEM:fig} the data
obtained using both methods are plotted. One can see that the \FEM\ results
also show a dependence on the sample resolution. As stated above, in the case
of the \FEM\ the \lr, \ir, and \hr\ consist of domains with 22048, 49670, and
982376~elements. The resolution dependency is also demonstrated in the inset of
the figure, where the permeability is shown as a function of resolution for a
Reynolds number of $\Re\approx 10^{-3}$. For simulations using $1\times10^6$
elements, there is a difference of
$\approx 2\%$ between \LB\ and \FEM\ results. This can be explained by several
factors: neither \FEM\ nor \LB\ results are fully converged with respect to the
required number of elements, but performing a large number of simulations with
substantially higher resolution is not feasible with the computational
resources available -- in particular since the final values are expected to
change by not more than a few percent. Since the \LB\ results are
systematically lower than the \FEM\ results, a possible explanation can be the
loss of accuracy of \LB\ due to the relatively small choice for the relaxation
time ($\tau/\dt=0.6$) as demonstrated in Fig.~\ref{k:hr:fig}. In any case, the
deviation is small and thus it can be concluded that the data in
Fig.~\ref{k:LB:FEM:fig} shows a qualitative and quantitative agreement when the
Reynolds number increases.
Fig.~\ref{k:LB:FEM:fig} shows that for both methods the value of the
permeability, beyond validity of Darcy's law, Eq.~\eqref{darcy:eq}, drops
consistently. Darcy's law is limited to the weak inertia regime, i.e. low
Reynolds numbers. To analyze the validity of the simulated data for high
Reynolds numbers, we plot $\us$ vs. $\dpl$ in Fig.~\ref{fit:fig} using the
\hr\ data and $\tau/\dt=0.6$ for \LBMRT. For both methods, at small velocity
values (small \Re), the flow follows Darcy's prediction (constant slope) and
Eq.~\eqref{darcy:eq} accurately fits the data. When the velocity increases the
measured permeability departs from Darcy's law and the best possible fit is
obtained using Eq.~\eqref{cubic:eq}, which confirms the existence of a
transition regime with cubic velocity dependence. Only for higher velocities
the flow enters Forchheimer's regime and the data can be fitted accurately by
Eq.~\eqref{forchheimer:eq}. The departure from the different flow regimes is
clearly seen in the bottom panel of Fig.~\ref{fit:fig}, where the relative
error of the fits is plotted versus \Re. It is calculated by
\begin{equation}
  \epsilon(w)=\frac{w_{\rm fit} - w_{\rm sim}}{w_{\rm sim}},
\end{equation}
where the indices $_{\rm sim}$ and $_{\rm fit}$ represent the results obtained by
simulation and from the fit, respectively. It can
clearly be seen from both datasets that Darcy's regime holds until $\Re
\approx \pwrr{0}$ and the inertia transition correction expression holds until $\Re
\approx \pwrr{1}$. The Forchheimer's regime describes the flow for higher \Re.

\section{Conclusions}
We presented an analysis of the calculation of the flow in porous
media in different flow regimes using the \LB\ method and compared our results
to \FEM\ simulations. Special attention has been given to discretization,
selection of a collision kernel and choice of parameters.
For the \LB\ method a simulation setup combining a constant pressure
at the inlet and outlet and an external acceleration acting inside the sample
has been proposed in order to reduce compressibility artifacts at high \Re.
However, at higher \Re\ a substantial number of lattice nodes shows flow
velocities beyond 20\% of the speed of sound and compressibility effects can
clearly be observed. In order to clarify the validity of these results we
compare our data to \FEM\ simulations and demonstrate a good quantitative
agreement for the full range of \Re\ studied.
Furthermore, the data show good agreement with theoretical predictions,
demonstrating that the range of \Re\ studied in this work is well
accessible for the \LB\ method and that compressibility effects only have
a minor influence. The results accurately predict three
different flow regimes. These are
Darcy's regime for $\Re \lesssim \pwrr{0}$, where the average pressure
gradient $\dpl$ and the surface velocity $\us$ obey a linear
relationship. Secondly, a transient regime $\pwrr{0} \lesssim \Re
\lesssim \pwrr{1}$, where the flow is modeled by Darcy's law plus an
inertia correction term represented by a cubic term on $\us$, see
Eq.~\eqref{cubic:eq}. Finally, for higher Reynolds numbers ($\Re \gtrsim
\pwrr{1}$) the porous media flow follows Forchheimer's prediction up to the limit of both simulation methods of Re $\approx$ 30.

\section{Acknowledgments}
Financial support by STW (STW-MuST program, project number 10120), NWO/STW
(VIDI project 10787 and VICI project 10828) and the FOM-Shell IPOG-II IPP is highly acknowledged.
Computing time was provided by the J\"ulich Supercomputing Centre (JSC)
and the Scientific Supercomputing Center Karlsruhe (SSCK).

%%% Bibliography %%%
%\bibliographystyle{bib/abbrv-unsrt} 
%\bibliography{bib/new_bib.bib}

\begin{thebibliography}{10}

\bibitem{Bear:1972}
J.~Bear.
\newblock {\em Dynamics of fluids in porous media}.
\newblock Elsevier (New York), 1972.

\bibitem{MPT_2009}
K.~Moutsopoulos, I.~Papaspyros, and V.~Tsihrintzis.
\newblock Experimental investigation of inertial flow processes in porous
  media.
\newblock {\em Journal of Hydrology}, 374(3-4):242--254, 2009.

\bibitem{bib:ferreol-rothman}
B.~Ferr\'{e}ol and D.~Rothman.
\newblock {L}attice-{B}oltzmann simulations of flow through {F}ontainebleau
  sandstone.
\newblock {\em Transp. Porous Media}, 20(1-2):3--20, 1995.

\bibitem{1990PhFl.2.2085C}
A.~Cancelliere, C.~Chang, E.~Foti, D.~H. Rothman, and S.~Succi.
\newblock The permeability of a random medium: Comparison of simulation with
  theory.
\newblock {\em Phys. Fluids.}, 2:2085--2088, 1990.

\bibitem{bib:qian-dhumieres-lallemand}
Y.~H. Qian, D.~{d'Humi\`{e}res}, and P.~Lallemand.
\newblock Lattice {BGK} models for {N}avier-{S}tokes equation.
\newblock {\em Europhys. Lett.}, 17(6):479--484, 1992.

\bibitem{PhysRevB.46.6080}
N.~Martys and E.~Garboczi.
\newblock Length scales relating the fluid permeability and electrical
  conductivity in random two-dimensional model porous media.
\newblock {\em Phys. Rev. B}, 46(10):6080--6090, 1992.

\bibitem{MAKHT02}
C.~Manwart, U.~Aaltosalmi, A.~Koponen, R.~Hilfer, and J.~Timonen.
\newblock Lattice-{B}oltzmann and finite-difference simulations for the
  permeability of three-dimensional porous media.
\newblock {\em Phys. Rev. E}, 66(1):016702, 2002.

\bibitem{bib:succi-01}
S.~Succi.
\newblock {\em The lattice {B}oltzmann equation for fluid dynamics and beyond}.
\newblock Oxford University Press, 2001.

\bibitem{bib:aharonov-rothman}
E.~Aharonov and D.~H. Rothman.
\newblock Non-{N}ewtonian flow (through porous media): a lattice {B}oltzmann
  method.
\newblock {\em Geophys. Research Lett.}, 20:679, 1993.

\bibitem{bib:auzeraisGRL96}
F.~Auzerais, J.~Dunsmuir, B.~Ferreol, N.~Martys, J.~Olson, T.~Ramakrishnan,
  D.~Rothman, and L.~Schwartz.
\newblock Transport in sandstone: a study based on three dimensional
  microtomography.
\newblock {\em Geophys. Research Lett.}, 23(7):705, 1996.

\bibitem{KL97}
D.~Koch and A.~Ladd.
\newblock Moderate {R}eynolds number flows through periodic and random arrays
  of aligned cylinders.
\newblock {\em J. Fluid Mech.}, 349:31--66, 1997.

\bibitem{HKL01}
R.~Hill, D.~Koch, and A.~Ladd.
\newblock The first effects of fluid inertia on flows in ordered and random
  arrays of spheres.
\newblock {\em J. Fluid Mech.}, 448:213--241, 2001.

\bibitem{bib:cf.CPaLLuCMi.2006}
C.~Pan, L.-S. Luo, and C.~T. Miller.
\newblock An evaluation of lattice {B}oltzmann schemes for porous medium flow
  simulation.
\newblock {\em Computers \& Fluids}, 35:898--909, 2006.

\bibitem{NZRHH10}
A.~Narv\'aez, T.~Zauner, F.~Raischel, R.~Hilfer, and J.~Harting.
\newblock Quantitative analysis of numerical estimates for the permeability of
  porous media from lattice-{B}oltzmann simulations.
\newblock {\em J. Stat. Mech.: Theor. Exp.}, 2010:P11026, 2010.

\bibitem{NH10}
A.~Narv\'aez and J.~Harting.
\newblock Evaluation of pressure boundary conditions for permeability
  calculations using the lattice-{B}oltzmann method.
\newblock {\em Adv. in Appl. Math. and Mech.}, 2:685--700, 2010.

\bibitem{Zabaras}
N.~Zabaras and D.~Samanta.
\newblock A stabilized volume-averaging finite element method for flow in
  porous and binary alloy solidification processes.
\newblock {\em Int. J. for Num. Meth. in Eng.}, 60(5):1--38, 2004.

\bibitem{girault}
V.~Girault and P.~Raviart.
\newblock {\em Finite elements approximation of the Navier-Stokes equations}.
\newblock Springer Series SCM, 1986.

\bibitem{thomasset}
F.~Thomasset.
\newblock {\em Implementation of finite element methods for Navier-Stokes
  equations}.
\newblock Springer-Verlag, New York, 1981.

\bibitem{Tezduyar}
T.~Tezduyar, M.~Behr, and J.~Liou.
\newblock A new strategy for finite element computations involving moving
  boundaries and interfaces-the deforming spatial-domain/space-time procedure:
  I. the concept and the preliminary numerical tests.
\newblock {\em Comput. Meth. Appl. Mech. Eng.}, 94:339--351, 1992.

\bibitem{Fletcher}
C.~Fletcher.
\newblock {\em Computational Techniques for Fluid Dynamics}.
\newblock Springer, 1991.

\bibitem{Turek}
S.~Turek.
\newblock {\em Efficient Solvers for Incompressible Flow Problems: An
  Algorithmic and Computational Approach}.
\newblock Springer, 1999.

\bibitem{kazemluding}
K.~Yazdchi and S.~Luding.
\newblock Towards unified drag laws for inertial flow through fibrous
  materials.
\newblock {\em Chemical Engineering Journal}, 207:35--48, 2012.

\bibitem{kazemluding2}
K.~Yazdchi, S.~Srivastava and S.~Luding.
\newblock Micro-macro relations for flow through random arrays of cylinders.
\newblock {\em Composites Part A}, 43:2007--2020, 2012.

\bibitem{bib:darcy}
H.~Darcy.
\newblock {\em Les fontaines publiques de la ville de Dijon}.
\newblock Dalmont, Paris, 1856.

\bibitem{Dullien:1992}
F.~Dullien.
\newblock {\em Porous Media: Fluid Transport and Pore Structure}.
\newblock Academic Press, San Diego, 2nd edition, 1992.

\bibitem{MA91}
C.~Mei and J.-L. Aurialt.
\newblock The effect of weak inertia on flow through a porous medium.
\newblock {\em J. Fluid Mech.}, 222:647--663, 1991.

\bibitem{Mass1984}
G.~Massarani.
\newblock {\em Problemas em sistemas particulados}.
\newblock Ed. Edgard Blucher Ltda, R\'{i}o de Janeiro, 1984.

\bibitem{Forchh01}
P.~Forchheimer.
\newblock {W}asserbewegung durch {B}oden.
\newblock {\em Zeit. Ver. Deut. Ing.}, 45:1781--1788, 1901.

\bibitem{ACAMS_1999}
J.~Andrade, U.~Costa, M.~Almeida, H.~Makse, and H.~Stanley.
\newblock Inertial effects on fluid flow through disordered porous media.
\newblock {\em Phys. Rev. Lett.}, 82(26):5249--5252, 1999.

\bibitem{SA99}
E.~Skjetne and J.~Auriault.
\newblock High-velocity laminar and turbulent flow in porous media.
\newblock {\em Transp. Porous Media}, 36:131--147, 1999.

\bibitem{AGO07}
J.-L. Auriault, C.~Geindreau, and L.~Org\'{e}as.
\newblock Upscaling {F}orchheimer law.
\newblock {\em Transp. Porous Media}, 70:213--229, 2007.

\bibitem{WL91}
J.-C. Wodie and T.~Levy.
\newblock Corection non lin\'{e}aire de la loi de {D}arcy.
\newblock {\em C.R. Acad. Sci. Paris}, 312:1573--161, 1991.

\bibitem{FGQ97}
M.~Firdauss, J.-L. Guermond, and P.~{Le Qu\'{e}r\'{e}}.
\newblock Nonlinear corrections to {D}arcy's law at low {R}eynolds numbers.
\newblock {\em J. Fluid Mech.}, 343:331--350, 1997.

\bibitem{bib:chen-chen-martinez-matthaeus}
S.~Chen, H.~Chen, D.~Mart\'{i}nez, and W.~H. Matthaeus.
\newblock Lattice {B}oltzmann model for simulation of magnetohydrodynamics.
\newblock {\em Phys. Rev. Lett.}, 67(27):3776--3779, 1991.

\bibitem{bib:bgk}
P.~L. Bhatnagar, E.~P. Gross, and M.~Krook.
\newblock Model for collision processes in gases. {I}. small amplitude
  processes in charged and neutral one-component systems.
\newblock {\em Phys. Rev.}, 94(3):511--525, 1954.

\bibitem{Sterling:1996:SAL:227647.227666}
J.~Sterling and S.~Chen.
\newblock Stability analysis of lattice {B}oltzmann methods.
\newblock {\em J. Comp. Phys.}, 123(1):196--206, 1996.

\bibitem{bib:qian}
Y.~H. Qian.
\newblock Fractional propagation and the elimination of staggered invariants in
  lattice-{BGK} models.
\newblock {\em Int. J. Mod. Phys. C}, 8(4):753--761, 1997.

\bibitem{SukopThorne2007}
M.~C. Sukop and D.~T. {Thorne~(Jr.)}.
\newblock {\em Lattice {B}oltzmann Modeling, An Introduction for Geoscientists
  and Engineers}.
\newblock Springer, second edition, 2007.

\bibitem{2002RSPTA.360..437D}
D.~{d'Humi\`{e}res}, I.~Ginzburg, M.~Krafczyk, P.~Lallemand, and L.-S. Luo.
\newblock {Multiple-relaxation-time lattice {B}oltzmann models in three
  dimensions}.
\newblock {\em Phil. Trans. R. Soc. Lond. A}, 360(1792):437--451, 2002.

\bibitem{2000PhRvE..61.6546L}
P.~Lallemand and L.-S. Luo.
\newblock Theory of the lattice {B}oltzmann method: dispersion, dissipation,
  isotropy, {G}alilean invariance, and stability.
\newblock {\em Phys. Rev. E}, 61(6):6546--6562, 2000.

\bibitem{el00178}
I.~Ginzburg, F.~Verhaeghe, and D.~{d'Humi\`{e}res}.
\newblock Two-relaxation-time lattice {B}oltzmann scheme: about
  parametrization, velocity, pressure and mixed boundary conditions.
\newblock {\em Comm. Comp. Phys.}, 3(2):427--478, 2008.

\bibitem{el00173}
I.~Ginzburg, F.~Verhaeghe, and D.~{d'Humi\`{e}res}.
\newblock Study of simple hydrodynamic solutions with the two-relaxation-times
  lattice {B}oltzmann scheme.
\newblock {\em Comm. Comp. Phys.}, 3(3):519--581, 2008.

\bibitem{bib:chapman-cowling}
S.~Chapman and T.~G. Cowling.
\newblock {\em The mathematical theory of non-uniform gases}.
\newblock Cambridge University Press, second edition, 1952.

\bibitem{Wolf05}
D.~A. {Wolf-Gladrow}.
\newblock {\em {L}attice-{G}as {C}ellular {A}utomata and lattice {B}oltzmann
  models}.
\newblock Springer, 2005.

\bibitem{Kandhai_1999}
D.~Kandhai, D.-E. Vidal, A.~Hoekstra, H.~Hoefsloot, P.~Iedema, and P.~Sloot.
\newblock Lattice-{B}oltzmann and finite element simulations of fluid flow in a
  smrx static mixer reactor.
\newblock {\em Int. J. Numer. Meth. Fluids}, 31:1019--1033, 1999.

\bibitem{saurabh_2012}
S.~Srivastava, K.~Yazdchi, and S.~Luding.
\newblock Meso-scale coupling of {FEM/DEM} for fluid-particle interactions.
\newblock {\em In preparation}, 2012.

\bibitem{Langtangen}
H.~Langtangen, K.-A. Mardal, and R.~Winther.
\newblock Numerical methods for incompressible viscous flow.
\newblock {\em Adv. in Water Res.}, 25(8-12):1125--1146, 2002.

\bibitem{bib:Yazdchia}
K.~Yazdchi, S.~Srivastava, and S.~Luding.
\newblock Microstructural effects on the permeability of periodic fibrous
  porous media.
\newblock {\em Int. J. of Multiphase Flow}, 37(8):956--966, 2011.

\bibitem{Guo02}
Z.~Guo and T.~Zhao.
\newblock Lattice {B}oltzmann model for incompressible flows through porous
  media.
\newblock {\em Phys. Rev. E}, 66:036304, 2002.

\bibitem{bib:pf.QZoXHe.1997}
Q.~Zou and X.~He.
\newblock On pressure and velocity boundary conditions for the lattice
  {B}oltzmann {BGK} model.
\newblock {\em Phys. Fluids.}, 9(6):1591--1598, 1997.

\bibitem{HH08b}
M.~Hecht and J.~Harting.
\newblock Implementation of on-site velocity boundary conditions for {D3Q19}
  lattice {B}oltzmann simulations.
\newblock {\em J. Stat. Mech.: Theor. Exp.}, 2010(01):P01018, 2010.

\end{thebibliography}

\end{document}